\documentclass[sigconf]{acmart}

\usepackage{multirow}
\usepackage{forest}
\usepackage{graphicx}
\usepackage{subcaption}
\usepackage{balance}

\def\BibTeX{{\rm B\kern-.05em{\sc i\kern-.025em b}\kern-.08emT\kern-.1667em\lower.7ex\hbox{E}\kern-.125emX}}
    
\setcopyright{acmcopyright} 
\begin{document}

\fancyhead{}

\title{Exploiting Temporal Relationships in Video Moment Localization with Natural Language}

\author{Songyang Zhang}
\email{szhang83@ur.rochester.edu}
\affiliation{%
  \institution{University of Rochester}
  \city{Rochester}
  \state{New York}
  \country{USA}
  \postcode{14623}
  }

\author{Jinsong Su}
\email{jssu@xmu.edu.cn}
\affiliation{%
  \institution{Xiamen University}
  \city{Xiamen}
  \state{Fujian}
  \country{China}
  }

\author{Jiebo Luo}
\email{jluo@cs.rochester.edu}
\affiliation{%
  \institution{University of Rochester}
  \city{Rochester}
  \state{New York}
  \country{USA}
  \postcode{14623}
  }

%
\begin{abstract}
We address the problem of video moment localization with natural language, i.e. localizing a video segment described by a natural language sentence. 
While most prior work focuses on grounding the query as a whole, temporal dependencies and reasoning between events within the text are not fully considered.
In this paper, we propose a novel Temporal Compositional Modular Network (TCMN) where 
a tree attention network first automatically decomposes a sentence into three descriptions with respect to the main event, context event and  temporal signal.
Two modules are then utilized to measure the visual similarity and location similarity between each segment and the decomposed descriptions.
Moreover, since the main event and context event may rely on different modalities (RGB or optical flow),
we use late fusion to form an ensemble of four models, where each model is independently trained by one combination of the visual input.
Experiments show that our model outperforms the state-of-the-art methods on the TEMPO dataset.
\end{abstract}

%
%


%

\copyrightyear{2019} 
\acmYear{2019} 
\acmConference[MM '19]{Proceedings of the 27th ACM International Conference on Multimedia}{October 21--25, 2019}{Nice, France}
\acmBooktitle{Proceedings of the 27th ACM International Conference on Multimedia (MM '19), October 21--25, 2019, Nice, France}
\acmPrice{15.00}
\acmDOI{10.1145/3343031.3350879}
\acmISBN{978-1-4503-6889-6/19/10}

%
    \maketitle

\section{Introduction}
\begin{figure}[t]
\includegraphics[width=0.5\textwidth]{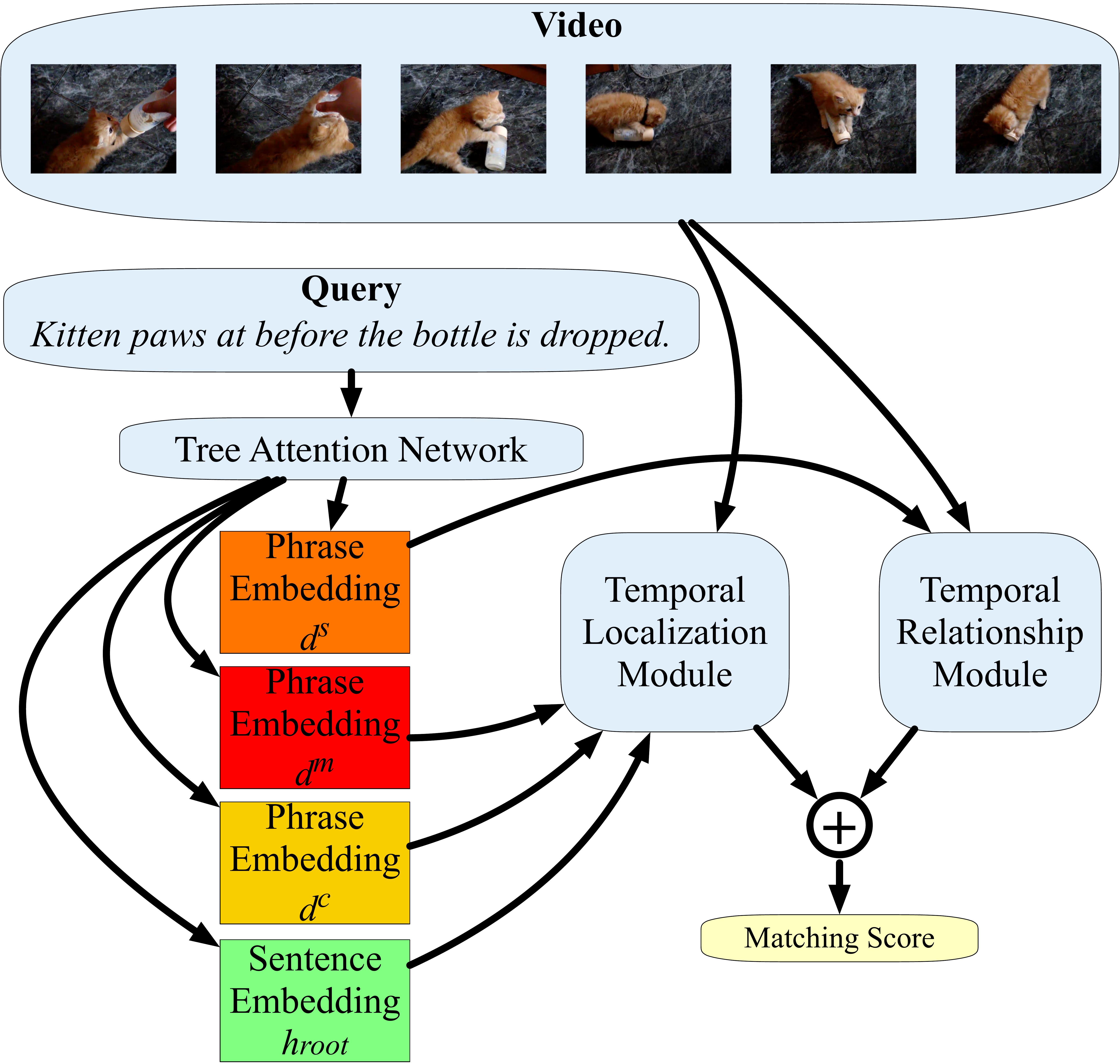}
	\caption{The pipeline of our proposed model}
	\label{fig:illustration}
\vspace{-4mm}
\end{figure}

Moment localization with temporal language aims to locate a segment in a video referred to by temporal language, which describes relationships between multiple events in a video.
It requires the model to be capable of localizing a single event and reasoning among multiple events.
In Figure~\ref{fig:illustration}, for example, the description \textit{kitten paws at before the bottle is dropped} is composed of a main event, \textit{kitten paws at the bottle}, a context event \textit{the bottle is dropped}, and their temporal ordering \textit{before}. 
Localizing a single event description has been explored by recent work~\cite{gao2017tall,hendricks17iccv,liu2018attentive,liu2018crossmodal,liu2018tmn,Ge_2019_WACV,chen2018temporally,chen2019localizing,xu2019multilevel,he2019read}, where most of them focus on elaborating the interaction between words and frames.
However, the description that involves multiple events has not been sufficiently investigated.

Therefore, in this work, we propose a temporal compositional modular network (TCMN) for moment localization with natural language, which exploits the temporal relationship hidden within the text to guide the network in localizing the corresponding segment.
There are three main novelties in our work.

\begin{figure}[t]
\centering
\begin{subfigure}[t]{0.3\textwidth}
	\includegraphics[width=\textwidth]{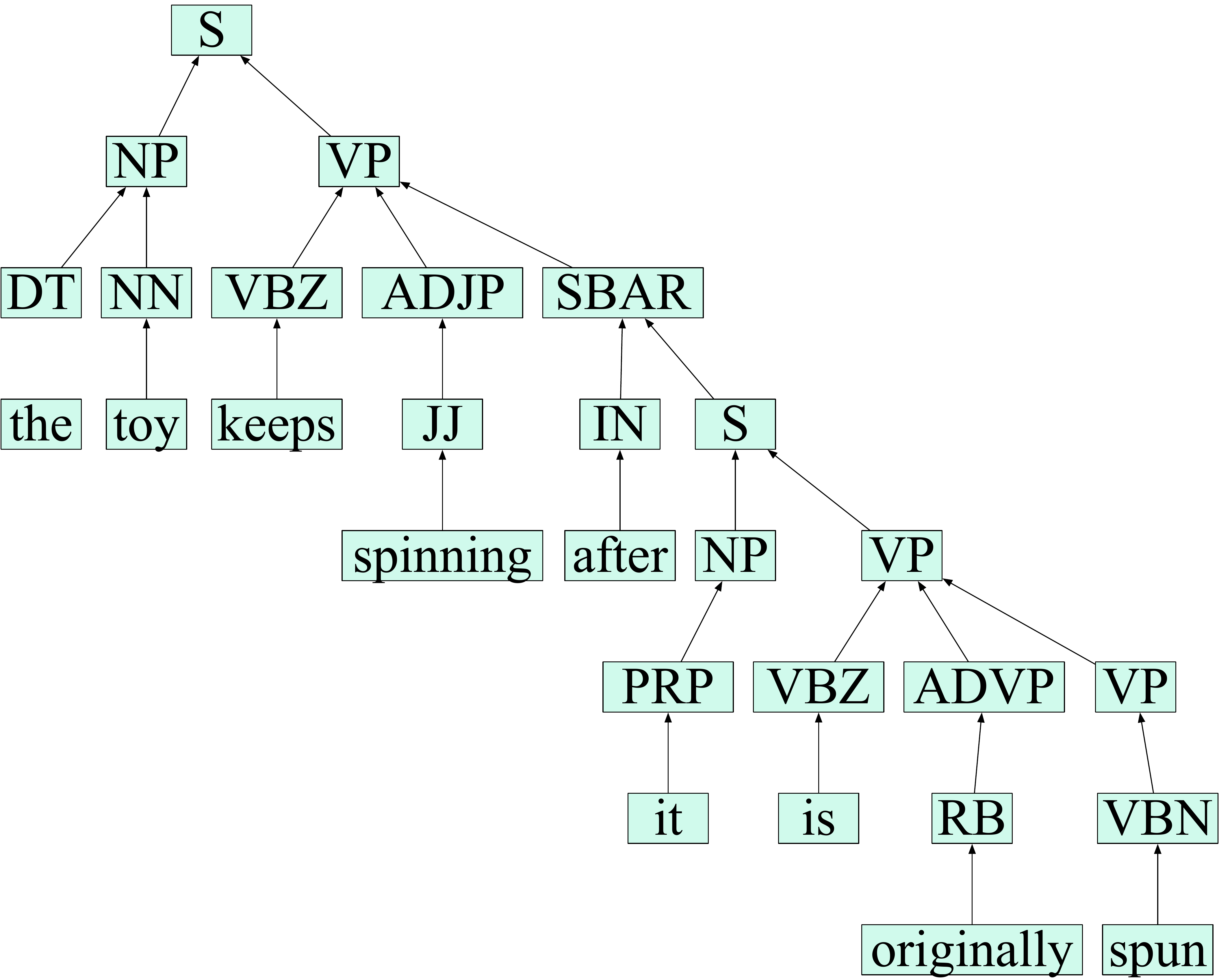}
	\caption{\textit{SBAR-TMP}}
\end{subfigure}
\hfill
\centering
\begin{subfigure}[t]{0.17\textwidth}
	\includegraphics[width=\textwidth]{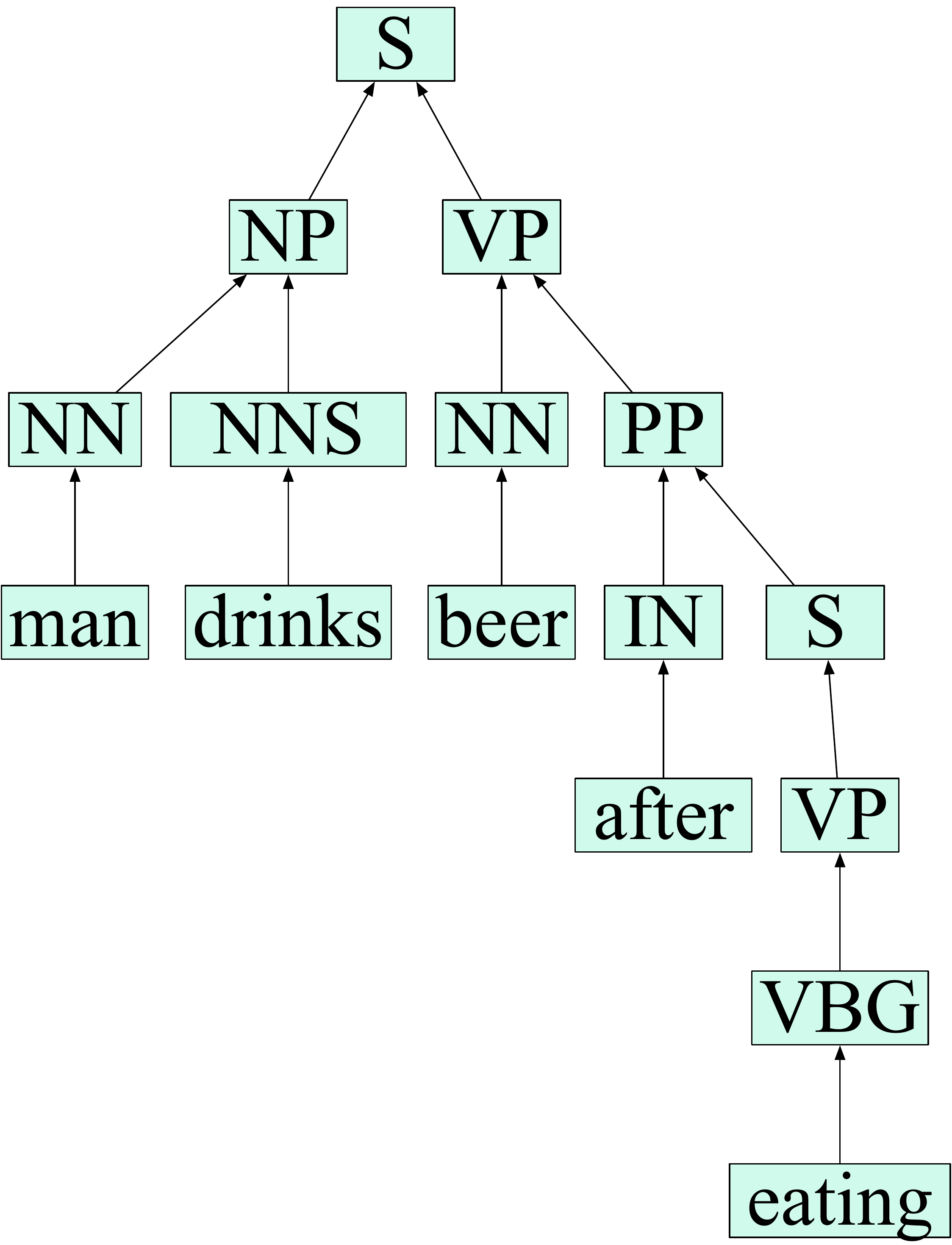}
	\caption{\textit{PP-TMP}}
\end{subfigure}
\caption{Two typical structures of temporal language.}
\label{fig:TMP}
\vspace{-6mm}
\end{figure}

First, a tree attention network is proposed to parse a given query through an attention mechanism.
A large portion of temporal language has a similar parsing structure~\cite{derczynski2017automatically}, as shown in Figure~\ref{fig:TMP}. 
Specifically, a typical temporal language parse tree is composed of three parts, namely the nodes for the main event, context event and temporal signal. 
Based on this observation, TCMN is designed to take the parsing tree of the query as language input. 
It then attends relevant nodes into three phrase embeddings, with respect to the main event, context event and temporal signal. 
Such a division allows our method to learn which segment is relevant to the query in detail.
Many efforts have been made in the vision and language area~\cite{liang2018focal,devlin2018bert,vaswani2017attention,xiang2018s3d}. 
The closest task with ours in this area is the reference expression comprehension, however, their language modeling is different from us.
To be specific, the entities for the reference expression are objects, while the entities for the temporal language are events.
The description of an object is often a single word or a single phrase, while the description for an event is often a single sentence, a single clause or a single phrase.
Therefore, we argue that when adapting the compositional modular network to video domain, applying attention over each word based on a sequential LSTM may not precisely parse the important cues for moment localization.

Second, two modules are designed for cross-modal matching from different perspectives.
The description for events is relevant to the  visual content, while the temporal signal is not. 
For example, \textit{before} describes the temporal ordering, which is irrelevant to the visual content of the corresponding event. 
When modeling such a relationship, introducing extra visual information may not benefit or even degenerate the reasoning.
Therefore, we use two modules to compute the matching score from different aspects, where the overall score is the sum of the two modules' output.
The temporal localization module measures the similarity between the query and the visual feature by utilizing the phrase embedding for the main event and context event,
while the relationship module measures the similarity between the phrase embeddings for the temporal signal and location feature.

Third, we use an ensemble model to solve the potential inconsistency of the visual modalities in this task.
The inconsistency here means that the events described in the searching query may rely on different visual modalities.
For example, given the sentence \textit{people are dancing before the darkest point}, where \textit{people are dancing} and \textit{the darkest point} are the descriptions for the main event and context event,  respectively.
Optical flow may be representative for human dancing, but it may be useless when describing darkness.
It is not reasonable to assume that both the main event and context event in a sentence rely on the same modality. 
In order to solve this problem, we form an ensemble of four models, where each model is independently trained by one combination of the visual input. Specifically, the visual feature for the main event and context event may be one of \{(RGB,RGB), (RGB,Flow), (Flow,RGB), (Flow,Flow)\}.
Compared to reference expression comprehension, localizing an image region in the spatial domain only involves the information on appearance from still frames, while localizing a video segment in the  temporal domain involves additional information on the motion between frames.
Therefore, the inconsistency of visual modalities is a unique problem in the video domain. Handling this problem is crucial for localizing a  video segment when the language involves temporal reasoning.

Our contributions are summarized as follows:
\begin{itemize}
    \item We propose a novel model called Temporal Compositional Modular Network that first learns to softly decompose a sentence into three descriptions with respect to the main event, context event and temporal signal, and then guides cross-modal feature matching by measuring the visual similarity and location similarity between each segment and the decomposed descriptions.

\item We further form an ensemble model to handle multiple events that may reflect on different visual modalities.

\item We achieve the state-of-the-art performance in the TEMPO dataset~\cite{hendricks18emnlp}, a diverse dataset for temporal reasoning in video and language. 
\end{itemize}

The rest of the paper is organized as follows. 
Section~\ref{sec:related work} reviews the related work. 
Section~\ref{sec:model} describes our proposed model.
We present the experimental results in Section~\ref{sec:experiment}, followed by the conclusion in Section~\ref{sec:conclusion}.

\section{Related Work}
\label{sec:related work}

\subsection{Temporal Signals}

Temporal signals are particular words or phrases that describe the temporal relation type, such as  \textit{before},\textit{after} and \textit{for the first time}, which provide context information for ordering~\cite{hitzeman1995semantic}.
Prior work~\cite{bree1993towards,bree1986temporal} has considered rule-based temporal inference by using temporal conjunctions and prepositions.
Schluter et al.~\cite{schluter2002temporal} list temporal signals in English expressions and compare their frequencies in British and US English.
Derczynski et al.~\cite{derczynski2017automatically} suggest that, a large portion of temporal signal expressions has a \textit{SBAR-TMP} or \textit{PP-TMP} subtree structure, where \textit{SBAR} is the mark for clause that is introduced by a (possibly empty) subordinating conjunction, \textit{PP} is the label for the prepositional phrase and the suffix \textit{-TMP} is a functional tag that indicates the existence of temporal adverbials.
This subtree usually begins with a temporal signal and also contains one of the signal's arguments, as shown in Figure~\ref{fig:TMP}.
It indicates that both the labels and word representations are strong indicators for dividing events and their relation within the text.
Different from previous work where the temporal signals are predefined, our framework can automatically learn the temporal signals from sentences in a data driven fashion.
In order to improve the performance of localizing the descriptions for the main event,  context event and  temporal signal, we use a tree attention network to consider each node.
Temporal ordering can be expressed linguistically in other ways, for example, tense can be used to describe the relation between other events~\cite{reichenbach2005tenses}.
Our work is different from this line of research, as we are primarily concerned about the temporal language tree structure and how the nodes refer to video segments.

\subsection{Moment Localization with Natural Language}

This task was recently introduced by Gao et al.~\cite{gao2017tall} and Hendricks et al.~\cite{hendricks17iccv}, which aims to localize the start and end time points within a long untrimmed video described by a sentence.
Gao et al.~\cite{gao2017tall} introduce a regression loss, while
Hendrick et al.~\cite{hendricks17iccv} simplify the problem by choosing from a set of pre-defined video segments.
Both models consider query-independent visual features as the context and encode the query as a whole.

Later on, there are three major directions for  handling this problem.
The first direction is to enhance the  representations of video segment features.
Liu et al.~\cite{liu2018attentive} design a memory attention model to emphasize the visual features mentioned in the query.
Xu et al.~\cite{xu2019multilevel} inject text features when generating clip proposals.
Zhang et al.~\cite{zhang2019man} introduce a graph convolution network to enhance the segment representations.

The second direction is to attend the most important words corresponding to the visual feature.
Liu et al.~\cite{liu2018crossmodal} devise a language-temporal attention model to adaptively identify the useful word information based on the temporal context.
Chen et al.~\cite{chen2018temporally} explore frame-by-word interactions between the video and language.
Meanwhile, they~\cite{chen2019localizing} also introduce a cross-gated mechanism for exploiting the fine-grained interactions.

The third direction is to introduce additional parsing information of the query.
Ge et al.~\cite{Ge_2019_WACV} additionally encode \textit{verb-obj} pairs in the query sentence and leverage an action classifier to dynamically compute the visual attention over the query and its context information.
Liu et al.~\cite{liu2018tmn} design a recursive neural network to dynamically fuse the visual feature and textual feature based on the query parsing result.


A common limitation of these previous works is that they assume the sentence is only related to one event, while the sentence for temporal language, where multiple events are involved, are not fully considered. 
Recently, Hendricks et al.~\cite{hendricks18emnlp} collect a new dataset to evaluate the model's performance with respect to the temporal language.
They propose to compute the similarity between the pairwise segment feature and the encoding for the full sentence.
However, using a single textual embedding to represent the entire sentence may be not precise enough. 
We argue that exploiting a detailed relationship between visual and textual information may further improve the performance.

\subsection{Reference Expression Comprehension}

Reference expression comprehension aims to localize a region within an image with a given referring expression.
The key for grounding the referring expression is the context information to distinguish the target from others.
For example, given the expression \textit{largest elephant standing behind baby elephant} and an image with region proposals, the model is required to identify the target elephant from others.

There are extensive efforts on the computation of the matching score between each region proposal and the given expression.
Hu et al.~\cite{hu2017modeling} propose the compositional modular network(CMN) composed of  three modules, which identify the subject, relationship and object, respectively.
Yu et al.~\cite{yu2018mattnet} build MattNet upon CMN, which further elaborates on the weight computing for each module and utilizes language-based attention and visual attention to focus on the  relevant regions.
Zhang et al.~\cite{zhang2018grounding} also build upon CMN and introduce variational inference to generate the context for each segment.
Liu et al.~\cite{liu2019improving}  build upon MattNet, and design a cross-model erasing mechanism to explore complementary cross-modal alignments.

Although these models are shown to be effective in their original tasks, simply extending them to video moment localization is not applicable because it does not account for the temporal information in videos.

\section{Model}
\label{sec:model}

Our proposed TCMN is composed of a tree attention network, a localization module and a relationship module.
Given a candidate video segment $v_i$ and a query $q$, we first use the tree attention network to perform a soft parsing of the given query into three components (one for the relationship module and two for the localization module) and map each to a phrase embedding.
Next, we use the relationship and localization modules to compute the similarity scores for segment $v_i$ to their respective textual embeddings.
Four models are trained independently, where each corresponds to one combination of RGB and optical flow input.
Finally, we use late fusion to combine all these scores and compute an overall matching score to  measure the similarity between $v_i$ and $q$.

\subsection{Tree Attention Network}

\begin{figure}[t]
\includegraphics[width=0.5\textwidth]{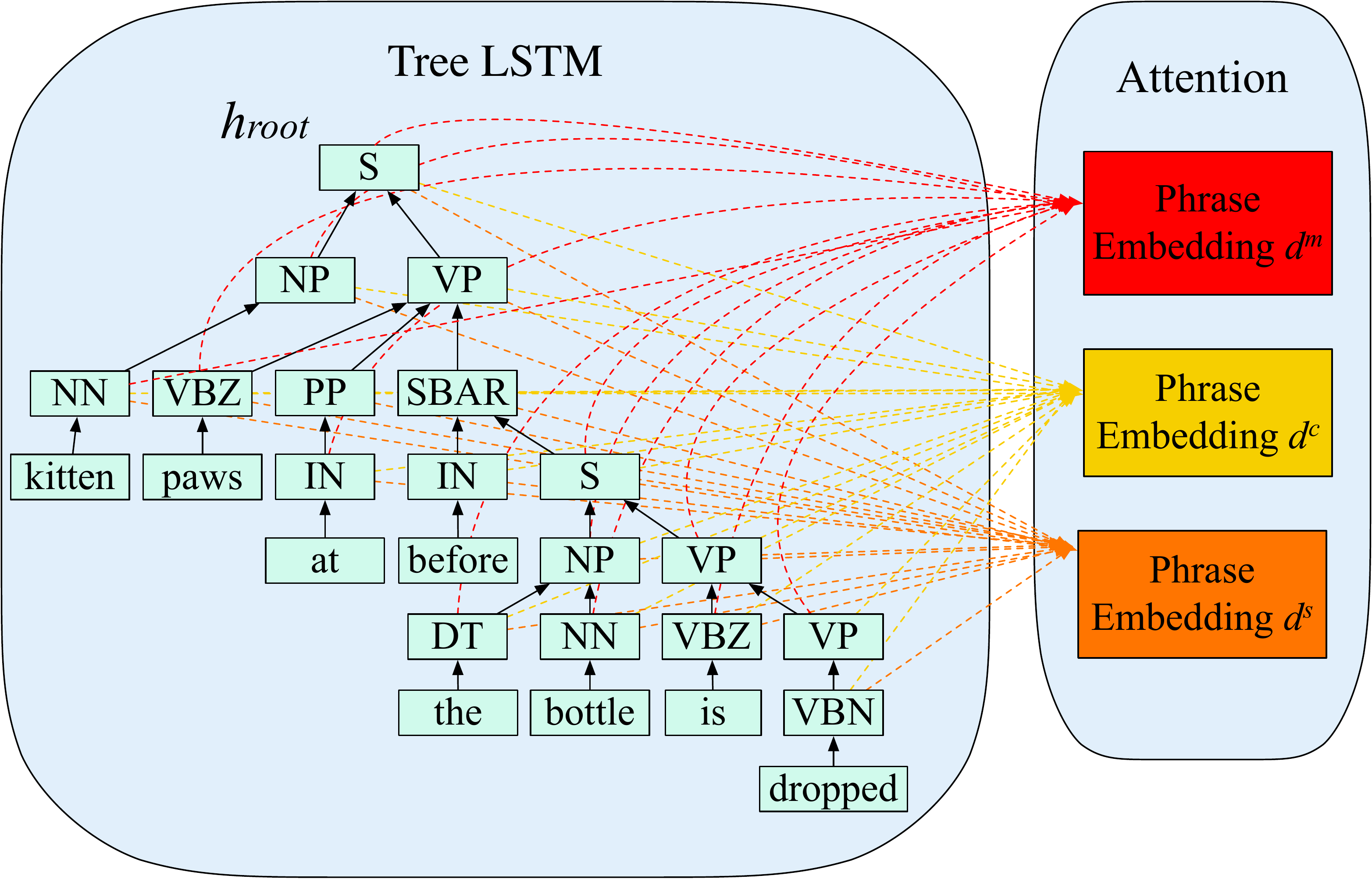}
	\caption{Illustration of the tree attention network.}
	\label{fig:treeattention}
\vspace{-2mm}
\end{figure}

Since most temporal languages share a similar structure after being parsed to a constituent tree, we encode such structured information into our language model.
The Tree-LSTM~\cite{tai2015improved} has been proposed to model the tree structure of sentences. 
In particular, a sentence is first parsed as a tree, where the root node indicates the representation of the full sentence while the intermediate nodes represent the phrases within the sentence.
Simply predefining a set of accepted labels and words to extract the corresponding event descriptions and temporal signals may not be general enough.
Instead, we design a tree attention network to utilize such structured information, as shown in Figure~\ref{fig:treeattention}.
Specifically, we use tree LSTM to extract the  feature for each node, as given in Equation~\ref{eq:treelstm}.

\begin{equation}
	\begin{aligned}
 		H &= TreeLSTM(\{x_i\}_{i=1,2,...,K}),\\
	\end{aligned}
	\label{eq:treelstm}
\end{equation}
where $x_i$ is the input word embeddings and $H=[h_1,h_2,...h_n]$ is the encoded feature for each node.
We label the encoded feature for the root node as $h_{root}$, which represents the whole sentence.

We then use an attention mechanism to compute the encoding of three phrases, which will be used later as the feature for the descriptions of the main event, context event and temporal signal, as given in Equation~\ref{eq:treeattention}.
\begin{equation}
	\begin{aligned}
		e_j &= embedding(l_j),\\
		\alpha^n &= softmax(w^n[h,e]+b^n),\\
		d^n &= \sum_j^N \alpha_j^nh_j,\\
	\end{aligned}
	\label{eq:treeattention}
\end{equation}
where $h_j$ and $l_j$ are the phrase encoding and the label's one hot encoding for node $j$.
$l_j$ is first embedded into a vector $e_j$.
The weight $w^n$ and bias $b^n$ are then used to compute the attention value $a^n$ on each node for each component $n$, where $n\in \{m,c,s\}$ ($m$ represents the main event, $c$ represents the context event, $s$ represents the temporal signals).
$[h,e]$ is the concatenation operation between $h$ and $e$. 
$N$ is the total number of nodes.
$d^n$ is the attended feature for component $n$.

\subsection{Temporal Localization Module}

\begin{figure}[t]
\includegraphics[width=0.45\textwidth]{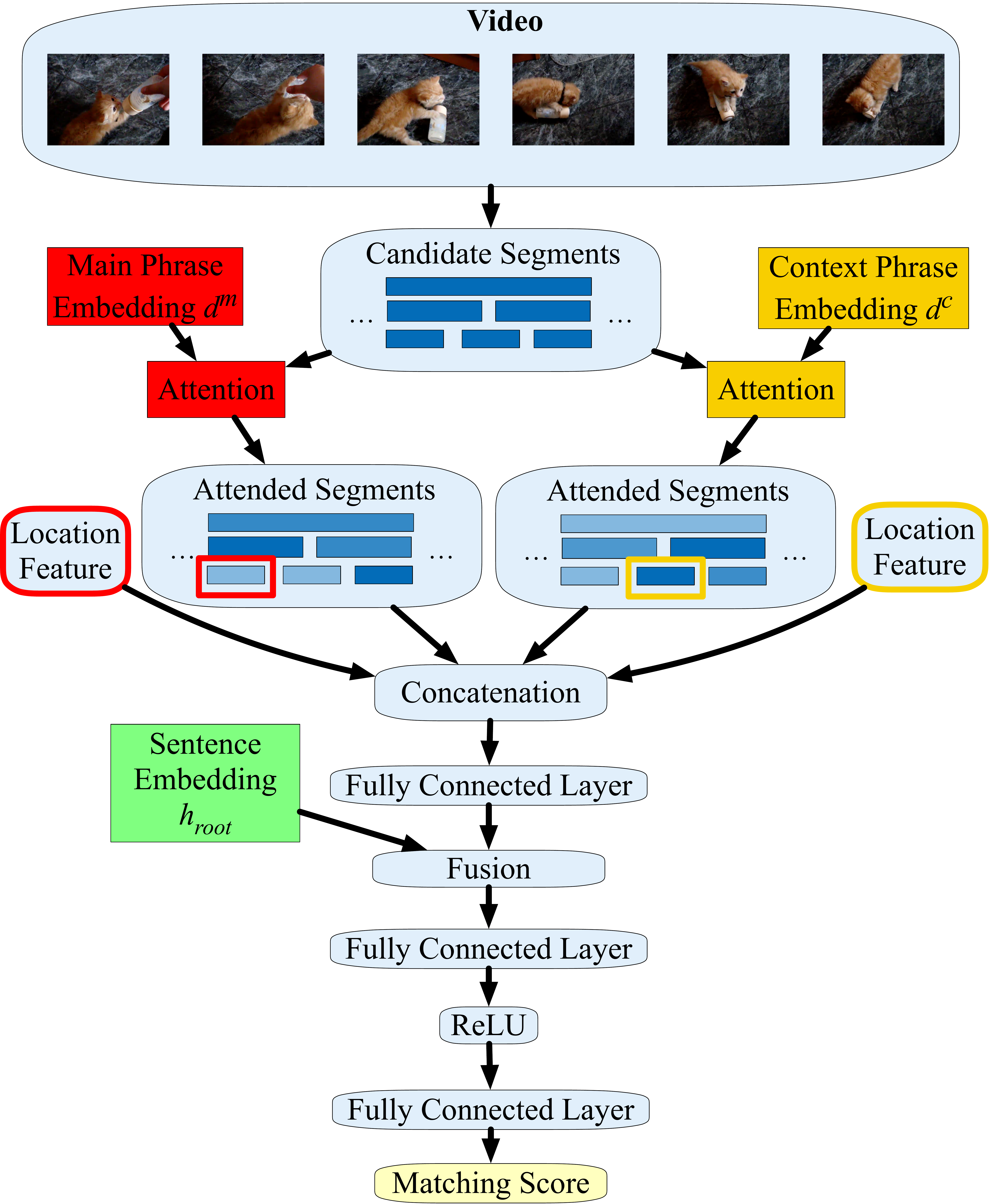}
	\caption{Illustration of the temporal localization module.}
	\label{fig:localizationmodule}
\vspace{-8mm}
\end{figure}

The temporal localization module is designed to compute the matching score between the given query and each video segment, as shown in Figure~\ref{fig:localizationmodule}.
Since temporal language usually involves multiple events, encoding the query holistically would ignore the cues for events within the text. 

We first use pretrained networks~\cite{simonyan2014very,wang2016temporal} to extract video frame features and then use multiple mean pooling layers with different strides to produce the segment feature $\{v_i\}_{i=1,2,...,P}$ where $P$ is the total number of segments.
Given the video segment features $\{v_i\}_{i=1,2,...,P}$ and the textual embeddings for the main event $d^{m}$, context event $d^{c}$ and full sentence $h_{root}$, we compute the matching scores as follows:
\begin{equation}
	\begin{aligned}
		&s^n_i = f^{n}(d^{n},v_i),\\
		&\alpha^n = softmax(s^n)\\
		&v_i^{n} = \alpha_i^nv_i,\\
		&t_{i} = [a_i,b_i],\\
		&s_{ij}^{loc} = f^{loc}(h_{root},[v_i^{m},t_{i},v_j^{c},t_{j}]),\\
	\end{aligned}
	\label{eq:locmodule}
\end{equation}
where $f^n$ first computes a score $s^n_i$ for video segment $v_i$ for component $n$, $n\in \{m,c\}$.
$s^n$ is then normalized by $softmax$ and output $\alpha^n$.
$v_i$ is weighted by $\alpha^n_i$ and output the video segment feature for component $n$.
$t_{i}$ is the start and end time points encoding for segment $i$, where $a$ and $b$ represent the start and end time points, respectively.
$f^{loc}$ computes the matching score $s_{ij}^{loc}$ between the segment pair $(i,j)$ and the full sentence encoding $h_{root}$.
$[v_i^{m},t_{i},v_j^{c},t_{j}]$ is the concatenated visual and location feature for the segment $i$ and $j$.
$f^m$, $f^c$ and $f^{loc}$ have the same structure, where both the textual feature and visual feature are first embedded to the same feature space with a fully connected layer for each feature.
The embedded features are then added up, normalized and fed into the fully connected layers to produce the corresponding scores.

\subsection{Temporal Relationship Module}

\begin{figure}[t]
\includegraphics[width=0.45\textwidth]{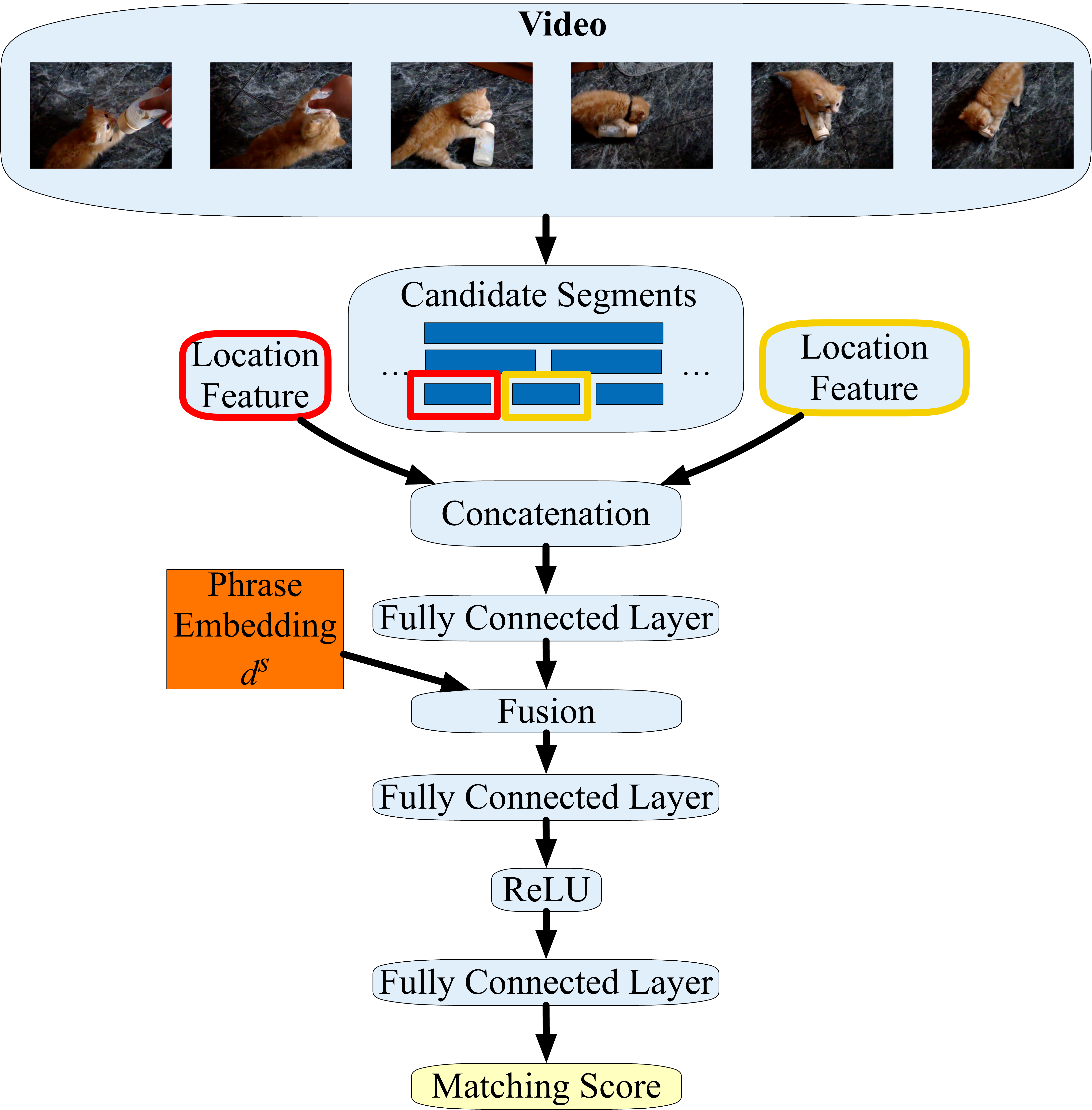}
	\caption{Illustration of the temporal relationship module.}
	\label{fig:relationshipmodule}
\vspace{-6mm}
\end{figure}

While the temporal localization module matches the visually relevant phrases with the video segments, there exist some visually irrelevant phrases that describe the temporal ordering between segments.
The temporal relationship module is designed to model such relationships, as shown in Figure~\ref{fig:relationshipmodule}.
Given the localization encoding $t_i$ and the prepared temporal signal encoding $d^{s}$, we calculate their matching score as follows:
\begin{equation}
	\begin{aligned}
		&s_{ij}^{rel} = f^{rel}(d^{s},[t_i,t_j]),\\
	\end{aligned}
	\label{eq:relmodule}
\end{equation}
where $f^{rel}$ computes the matching score $s_{ij}^{rel}$ between the video segment pair $(i,j)$ and the temporal signal encoding $d^{s}$. 
$[t_i,t_j]$ is the concatenated location encoding for the segment $i$ and $j$.
The structure of $f^{rel}$ follows the same design of $f^m$, $f^c$ and $f^{loc}$, where the only difference is the input. 

\subsection{Loss Function}

We use margin ranking loss as our loss function.
Two losses are utilized, where $L^{m}$ measures the loss for the main event and $L^{c}$ measures the loss for the context event, which are defined as follows:
\begin{equation}
	\begin{aligned}
		&s_{ij} = s_{ij}^{loc}+s_{ij}^{rel},\\
		&L^{m} = \frac{1}{N}\sum_{i\in N}max(0,\max_j(s_{ij})-\max_k(s_{pk})+M^{m}),\\
		&L^{c} = \frac{1}{N}\sum_{i\in N}max(0,s_{pi}-s_{pq}+M^{c}),\\
		&L = L^{m}+\lambda L^{c},\\
	\end{aligned}
	\label{eq:loss}
\end{equation}
where $p$ and $q$ are the indices of the ground truth segments for the main event and context event, respectively. $M^{m}$ and $M^{c}$ are the margins for the two losses, $N$ is the set of all possible segments, and $\lambda$ is a constant parameter to weigh the main event loss and context event loss.
For the queries that only involve one event, where the context ground truth is not available, we use the entire video as the context instead.


\subsection{Ensemble of Models}

Since different events may reflect on different visual modalities (RGB or optical flow), we make an ensemble of four models to predict the final score, where each model corresponds to one combination of event and visual feature pairs.

Specifically, there are two kinds of event, the main event and the context event.
For each event, the corresponding video segment feature comes from either RGB or optical flow.
Therefore, there are four combinations of visual features and each of them is fed to an independent model for training.
Late fusion is then employed to combine the four models to obtain the final scores.
Let $V$ be the set of all combinations of visual features, which is \{$(RGB,RGB)$, $(RGB,Flow)$, $(Flow,RGB)$, $(Flow,Flow)$\}, late fusion is defined as:
\begin{equation}
	\begin{aligned}
		&s_{ij} = \sum_{n\in V}\gamma^vs_{ij}^n,\\
		&\sum_{n\in V}\gamma_{n} = 1,\\
	\end{aligned}
	\label{eq:loss}
\end{equation}
where $s^n_{ij}$ is the matching score for the visual feature pair $n$, $\gamma_{n}$ is the weight for score $s_{ij}^n$ and the total sum of $\gamma$ is equal to one.

\subsection{Implementation Details}
We follow the same setting of computing the visual feature and textual feature in ~\cite{hendricks18emnlp}, where the RGB features are extracted from VGG~\cite{simonyan2014very} fc7 layer, optical flow features are extracted from the penultimate layer ~\cite{wang2016temporal} and the 300-d Glove feature~\cite{pennington2014glove} pretrained on Common Crawl (42 billion tokens) are used as the word embedding.
The segment visual feature are obtained by average-pooling over frames to produce a 4096-d vector (RGB) or a 1024-d (optical flow) vector.
Berkeley Neural Parser~\cite{kitaev2018selfattentive} is used to obtain the parse tree.
During our training, we use Adam~\cite{kingma2014adam} to optimize the network with a learning rate of $0.001$ and a  weight decay of $1e-8$. 
Both margins in the related loss functions are set to $0.1$ and $\lambda$ is set to $1$ in all experiments.

\section{Experiments}
\label{sec:experiment}

\begin{table*}[t]
	\begin{center}
		\begin{tabular}{|c|c|c|c|c|c|c|c|c|c|c|c|}
			\hline
			\multirow{2}*{Method} & \multicolumn{2}{c|}{DiDeMo} & \multicolumn{2}{c|}{Before} & \multicolumn{2}{c|}{After} & \multicolumn{2}{c|}{Then} & \multicolumn{3}{c|}{Average}\\
			\cline{2-12}
			 & R@1 & mIoU & R@1 & mIoU & R@1 & mIoU & R@1 & mIoU & R@1 & R@5 & mIoU \\
			\hline
			Frequeny Prior & $10.71$ & $20.67$ & $17.85$ & $24.22$ & $22.42$ & $25.76$ & $0.00$  & $24.73$ & $12.74$ & $52.58$ & $23.84$ \\
			TMN~\cite{liu2018tmn} & $20.19$ & $33.18$ & $22.24$ & $27.25$ & $20.14$ & $28.49$ & $9.23$  & $36.45$ & $17.95$ & $61.27$ & $31.34$ \\
			TALL+TEF~\cite{gao2017tall} & $20.95$ & $32.09$ & $27.13$ & $32.41$ & $26.30$ & $34.27$ & $4.84$ & $36.75$ & $19.80$ & $64.66$ & $33.88$ \\
			MCN+TEF~\cite{hendricks17iccv} & $24.85$ & $37.92$ & $32.28$ & $38.67$ & $26.08$ & $35.44$ & $25.07$  & $53.94$ & $27.07$ & $73.36$ & $41.49$ \\
			MLLC+conTEF~\cite{hendricks18emnlp} & $27.46$ & $\mathbf{41.20}$ & $35.31$ & $41.81$ & $29.38$ & $38.90$ & $26.83$ & $54.97$ & $29.74$ & $76.76$ & $44.22$ \\
			\hline
			Ensemble TCMN & $\mathbf{28.90}$ & $41.03$ & $\mathbf{37.68}$ & $\mathbf{44.78}$ & $\mathbf{32.61}$ & $\mathbf{42.77}$ & $\mathbf{31.16}$ & $\mathbf{55.46}$ & $\mathbf{32.85}$ & $\mathbf{78.73}$ & $\mathbf{46.01}$  \\
			\hline
		\end{tabular}
	\end{center}
	\caption{Performance comparison  with the  state-of-art approaches on TEMPO-TL~\cite{hendricks18emnlp}. TMN~\cite{liu2018tmn} is based on our implementation and other baselines are reported in~\cite{hendricks18emnlp}.}
	\label{tab:TEMPO_TL}
\vspace{-6mm}
\end{table*}

\begin{table*}[t]
	\begin{center}
		\begin{tabular}{|c|c|c|c|c|c|c|c|c|c|c|c|c|c|}
			\hline
			\multirow{2}*{Method} & \multicolumn{2}{c|}{DiDeMo} & \multicolumn{2}{c|}{Before} & \multicolumn{2}{c|}{After} & \multicolumn{2}{c|}{Then} & \multicolumn{2}{c|}{While} & \multicolumn{3}{c|}{Average}\\
			\cline{2-14}
			 & R@1 & mIoU & R@1 & mIoU & R@1 & mIoU & R@1 & mIoU & R@1 & mIoU & R@1 & R@5 & mIoU \\
			\hline
			Frequeny Prior & $19.43$ & $25.44$ & $29.31$ & $51.92$ & $0.00$ & $0.00$ & $0.00$  & $7.84$ & $4.74$ & $12.27$ & $10.69$ & $37.56$ & $19.50$ \\
			TMN~\cite{liu2018tmn} & $20.74$ & $35.29$ & $10.44$ & $21.07$ & $09.78$ & $22.83$ & $3.70$ & $29.81$ & $9.79$  & $31.17$ & $10.89$ & $50.40$ & $28.03$ \\
			TALL+TEF~\cite{gao2017tall} & $21.79$ & $33.55$ & $25.91$ & $49.26$ & $14.43$ & $32.62$ & $2.52$ & $31.13$ & $8.1$ & $28.14$ & $14.55$ & $60.69$ & $34.94$ \\
			MCN+TEF~\cite{hendricks17iccv} & $26.07$ & $39.92$ & $26.79$ & $51.40$ & $14.93$ & $34.28$ & $18.55$  & $47.92$ & $10.70$ & $35.47$ & $19.40$ & $70.88$ & $41.80$ \\
			MLLC+conTEF~\cite{hendricks18emnlp} & $27.38$ & $\mathbf{42.45}$ & $32.33$ & $56.91$ & $14.43$ & $37.33$ & $19.58$ & $50.39$ & $10.39$ & $35.95$ & $20.82$ & $71.68$ & $44.57$ \\
			\hline
			Ensemble TCMN & $\mathbf{28.77}$ & $42.37$ & $\mathbf{35.47}$ & $\mathbf{59.28}$ & $\mathbf{17.91}$ & $\mathbf{40.79}$ & $\mathbf{20.47}$ & $\mathbf{50.78}$ & $\mathbf{18.81}$ & $\mathbf{42.95}$ & $\mathbf{24.29}$ & $\mathbf{76.98}$ & $\mathbf{47.24}$ \\
			\hline
		\end{tabular}
	\end{center}
	\caption{Performance comparison with the  state-of-art approaches on TEMPO-HL~\cite{hendricks18emnlp}. TMN~\cite{liu2018tmn} is based on our implementation and other baselines are reported in~\cite{hendricks18emnlp}.}
	\label{tab:TEMPO_HL}
\vspace{-6mm}
\end{table*}

\subsection{Dataset Description}

We conduct experiments on the TEMPO dataset~\cite{hendricks18emnlp} to evaluate our method, which is designed for temporal reasoning in video and language.
This dataset is collected based on the DiDeMo dataset~\cite{hendricks17iccv}, where sentences only describe one event in videos. 
The TEMPO dataset makes further extensions on the language descriptions that involve multiple events, while keeping its videos the same.
The extended language expressions are collected based on four commonly used temporal words, \textit{before}, \textit{after}, \textit{while} and \textit{then}.
Simple sentences that come from DiDeMo are also included in their dataset.
There are two parts in TEMPO, TEMPO - (Template Language) TL, which is constructed by the original DiDeMo sentences with language templates and TEMPO - (Human Language) HL, which consists of pure human annotations.

\subsection{Quantitative Results}

\subsubsection{Evaluation Metric}
We use the same evaluation metric as defined in TEMPO~\cite{hendricks18emnlp}.
There are four categories in TEMPO-TL, \textit{DiDeMo}, \textit{before}, \textit{after} and \textit{then} and five categories in TEMPO-HL with an additional category \textit{while}. 
The sentences belonging to \textit{DiDeMo} are almost all simple sentences, each of which only describes one single event.
All the remaining sentences are categorized based on the keywords appeared in them.
For each category, the rank at one (R@1), rank at five (R@5), and mean intersection over union (mIoU) are computed.
The average value for each type of metric among the categories is also computed as an overall evaluation for the performance on temporal language.

\subsubsection{Results on TEMPO-TL}

\begin{table}[t]
	\begin{center}
		\begin{tabular}{|c|c|c|c|}
			\hline
			Model & Rank@1 & Rank@5 & mean IoU \\
			\hline
			MCN+TEF w/ aug~\cite{hendricks17iccv} & $24.85$ & $-$ & $37.92$ \\
			MCN+TEF w/o aug~\cite{hendricks17iccv} & $27.65$ & $-$ & $41.91$ \\
			TMN w/ aug~\cite{liu2018tmn} & $20.19$ & $76.08$ & $33.18$ \\
			TMN w/o aug~\cite{liu2018tmn} & $22.92$ & $76.08$ & $35.17$ \\
			TGN~\cite{chen2018temporally} & $28.23$ & $79.26$ & $\mathbf{42.97}$ \\
			MAN~\cite{zhang2019man} & $27.02$ & $\mathbf{81.70}$ & $41.16$ \\
			MLLC+conTEF~\cite{hendricks18emnlp} & $27.38$ & $-$ & $42.45$ \\
    		\hline
    		Ensemble TCMN & $\mathbf{28.90}$ & $79.00$ & $41.03$ \\
			\hline
		\end{tabular}
	\end{center}
	\caption{Performance comparison on DiDeMo~\cite{hendricks17iccv}. \textit{w/ aug} and \textit{w/o aug} represent with and without data augmentation.}
	\label{tab:DiDeMoResult}
\vspace{-6mm}
\end{table}

Sentences in TEMPO-TL are generated by several predefined templates, two for \textit{before} (\textit{A before B} and \textit{Before A, B}), two for \textit{after}  (\textit{A after B} and \textit{After A, B}) and one for \textit{then} (\textit{A then B}), where A, B are two simple sentences coming from DiDeMo.

We first compare our model with MLLC~\cite{hendricks18emnlp} on complex sentences in Table~\ref{tab:TEMPO_TL}.
We achieve superior performance in all sentence types compared to MLLC.
The only exception is the mIoU in the \textit{DiDeMo} category, which is only slightly lower.

Next, we compare the models designed for simple sentences, including TALL~\cite{gao2017tall}, MCN~\cite{hendricks17iccv} and TMN~\cite{liu2018tmn}, with the models designed for complex sentences, which include MLLC~\cite{hendricks18emnlp} and ours.
Since this dataset is generated by templates, this strategy can also be regarded as a method for data augmentation, which aims to enhance the model's capability of handling complex sentences.
For the models designed for simple sentences, applying such data augmentation may not benefit and sometimes degenerate the performance on simple sentences, as shown in Table~\ref{tab:DiDeMoResult}.
For example, with the data augmentation, TMN~\cite{liu2018tmn} and MCN~\cite{hendricks17iccv} perform worse on DiDeMo compared to their models without data augmentation. 
For models designed for complex sentences, we observe that they still achieve comparative results on the DiDeMo dataset,  meaning this type of models is compatible with both simple sentences and complex sentences.
The underlying reason is that directly applying such data augmentation is not suitable for those models designed for simple sentences.

In addition, although our method is not designed for simple sentences, it achieves comparative results on DiDeMo compared to the state-of-art methods~\cite{zhang2019man,chen2018temporally}.
Since these methods focus on the interaction between words and frames, which is orthogonal to our research, we believe that combining these two types of research may make further improvement.

\subsubsection{Results on TEMPO-HL}

Table~\ref{tab:TEMPO_HL} compares the performance of our ensemble TCMN with the prior work~\cite{liu2018tmn,gao2017tall,hendricks17iccv,hendricks18emnlp}.
TEMPO-HL is more challenging than TEMPO-TL as suggested by the worse performance reported on TEMPO-HL.
The reason is that human annotated sentences often involve anaphora, omission and so on,  making the sentence composition even more complex.
Similar to TEMPO-TL, we also observe that the performance on the  models~\cite{liu2018tmn,hendricks17iccv}  designed for single events drops.

Ensemble TCMN exhibits the best performance across all the metrics of the complex sentence and comparative results in simple sentences.
Specifically, the performance improvement for \textit{while} is larger than it for \textit{then}, while the improvements for \textit{after} and \textit{before} is in between.
One possibility is that different temporal signals have different difficulties for localization.
Considering a hard case, where the segments A and B are intersected, let $b_1$ and $b_2$ be the boundaries of their intersection. In this case, their visual features are similar and thus hard to differentiate.
The prediction of \textit{A then B} is the union of A and B, which does not require an accurate identification of $b_1$ and $b_2$.
However, the prediction of \textit{A while B} is the intersection between A and B, which requires the model to precisely identify $b_1$ and $b_2$.
Meanwhile, \textit{A before B} and \textit{A after B} only require either $b_1$ or $b_2$ to be accurate.


\subsection{Attention Visualization}

\begin{figure*}[t]
\centering
\begin{subfigure}[b]{\textwidth}
	\includegraphics[width=\textwidth]{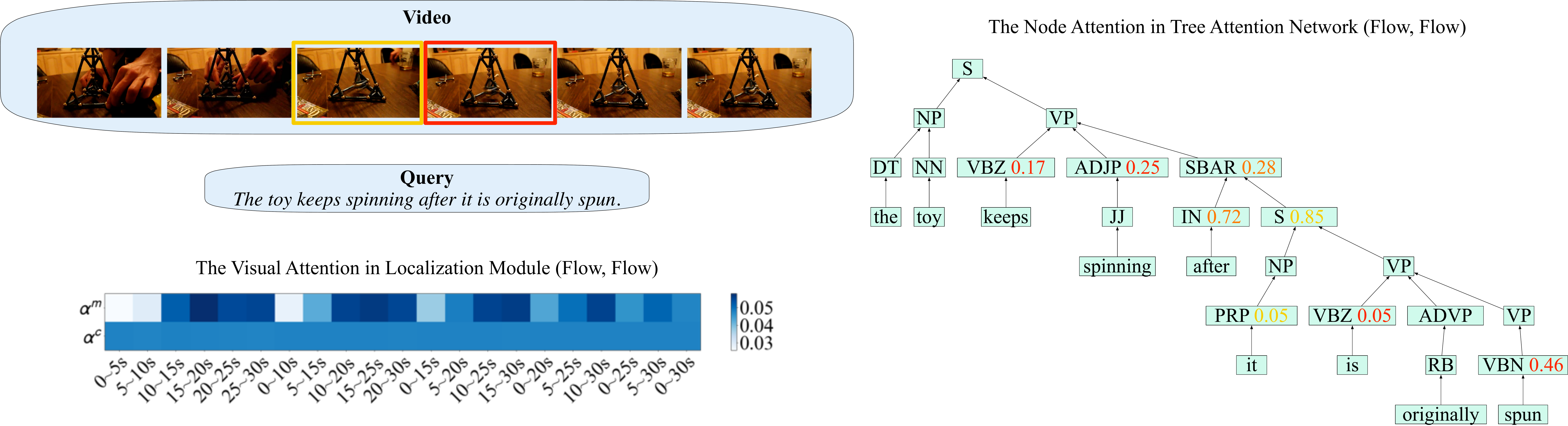}
	\caption{}
\end{subfigure}
\vspace{1em}
\begin{subfigure}[b]{\textwidth}
\includegraphics[width=\textwidth]{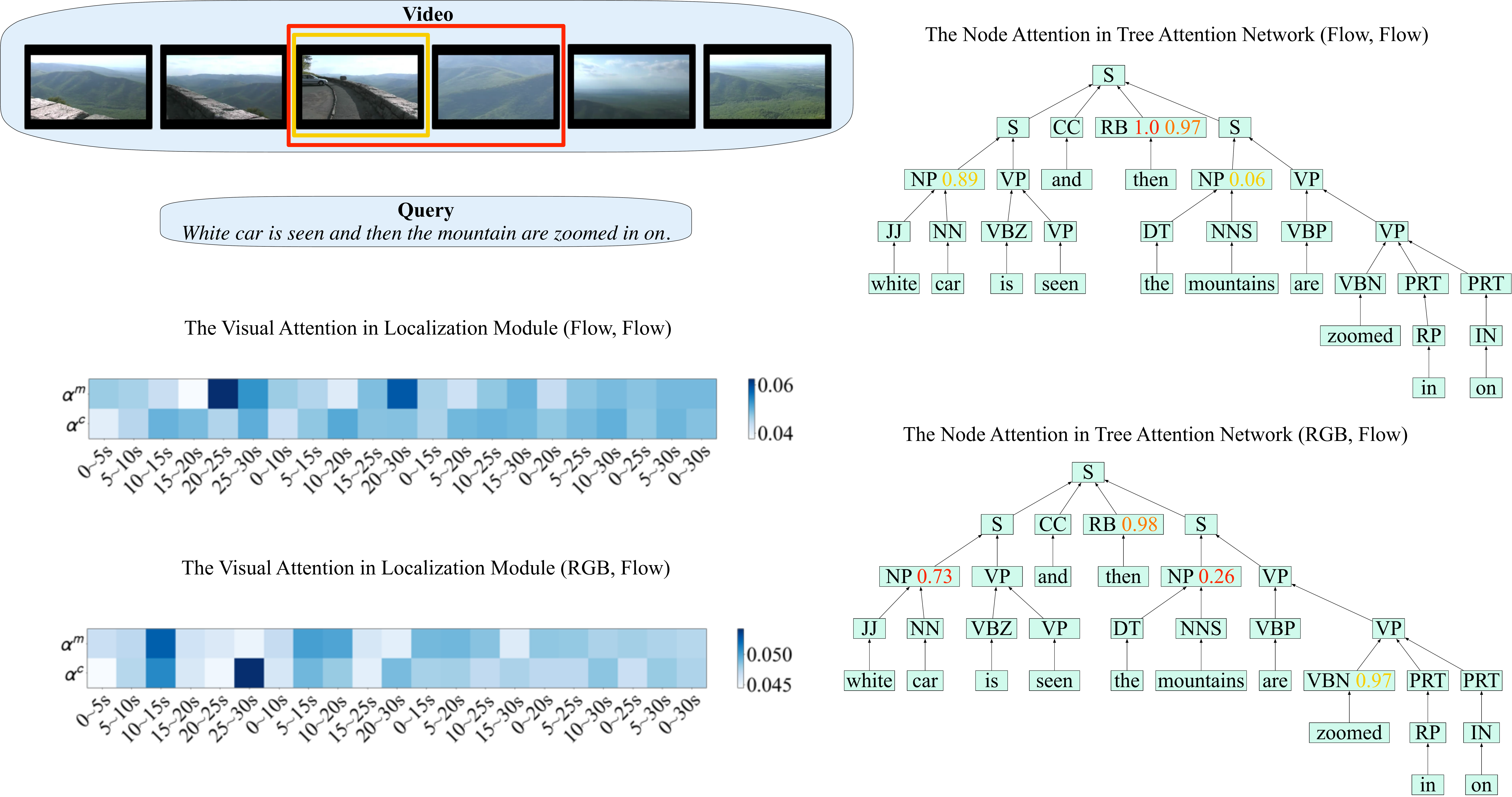}
	\caption{}
\end{subfigure}
	\caption{Visualization of attention values in TCMN. Node attention values that are less than $\mathbf{0.05}$ are not shown here. The red and yellow bounding boxes mark the ground truth segment corresponding to the full description and context description respectively. Each photo represents a five second video segment. The red, orange and yellow numbers represent the attention score for main event, temporal signal and context event for each tree node.}
	\label{fig:attentionvisualization}
\vspace{-2mm}
\end{figure*}
We visualize the visual attention values in the localization module and node attention values in the tree attention network in Figure~\ref{fig:attentionvisualization}. 

In Figure~\ref{fig:attentionvisualization} (a), we show an example that is correctly localized by the $(Flow, Flow)$ stream. 
Given a query \textit{the toy keeps spinning after it is originally spun}, the tree attention network computes three sets of attention values, where red, yellow and orange numbers represent the attention values for the main event, context event and temporal signal, respectively.
Different attentions focus on different nodes:
the attention for the main event concentrates on words \textit{keeps}, \textit{spining} and \textit{spun}, the attention for the context event highlights the \textit{SBAR}'s child sentence, and the attention for the temporal signal focuses on the word \textit{after}.
Moreover, from the visual attention visualization in the localization module, given the attended phrase encoding,  visual attention can diminish the influence of irrelevant segments, as shown in row $\alpha^m$, where the segments that are within 0\textasciitilde10s have lower scores.

In Figure~\ref{fig:attentionvisualization} (b), we show another example that is correctly localized by the $(RGB, Flow)$ stream but mistakenly grounded by the  $(Flow, Flow)$ stream. 
Without color information, even though the tree attention network in the $(Flow, Flow)$ stream attends the phrase \textit{white car} correctly, the visual attention can hardly localize the segment corresponding to it.
On the contrary, the visual attention in the $(RGB, Flow)$ stream can correctly localize the segment that has white car due to the RGB modality.

\subsection{Discussions}

\begin{table*}[t]
	\begin{center}
		\begin{tabular}{|c|c|c|c|c|c|c|c|c|c|c|c|c|c|c|}
			\hline
			\multirow{2}*{Method} & \multicolumn{2}{c|}{DiDeMo} & \multicolumn{2}{c|}{Before} & \multicolumn{2}{c|}{After} & \multicolumn{2}{c|}{Then} & \multicolumn{2}{c|}{While} & \multicolumn{3}{c|}{Average}\\
			\cline{2-14}
			 & R@1 & mIoU & R@1 & mIoU & R@1 & mIoU & R@1 & mIoU & R@1 & mIoU & R@1 & R@5 & mIoU \\
			\hline
			MLLC+conTEF~\cite{hendricks18emnlp} & $27.38$ & $42.45$ & $32.33$ & $56.91$ & $14.43$ & $37.33$ & $19.58$ & $50.39$ & $10.39$ & $35.95$ & $20.82$ & $71.68$ & $44.57$\\
			Ensemble MLLC+conTEF & $27.78$ & $42.91$ & $31.95$ & $56.72$ & $14.43$ & $37.64$ & $18.84$ & $50.33$ & $11.47$ & $36.48$ & $20.89$ & $73.05$ & $44.82$ \\
			TCMN w/ Two Streams & $27.56$ & $40.66$ & $35.89$ & $58.82$ & $\mathbf{22.45}$ & $\mathbf{45.21}$ & $17.94$ & $47.32$ & $17.66$ & $38.21$ & $24.30$ & $75.81$ & $46.05$ \\
			TCMN w/ REC Att & $\mathbf{29.35}$ & $\mathbf{43.75}$ & $31.82$ & $55.17$ & $17.58$ & $40.30$ & $19.58$ & $49.92$ & $13.15$ & $36.29$ & $22.30$ & $76.08$ & $45.08$ \\
			TCMN w/o Any Att & $27.61$ & $42.00$ & $31.82$ & $55.48$ & $17.58$ & $38.98$ & $15.28$ & $49.18$ & $13.30$ & $35.26$ & $21.12$ & $70.96$ & $44.18$ \\
			TCMN w/o Loc Att & $28.67$ & $42.98$ & $34.09$ & $59.18$ & $18.41$ & $41.28$ & $20.18$ & $50.75$ & $\mathbf{19.27}$ & $40.78$ & $24.12$ & $75.95$ & $46.99$ \\
			\hline
			TCMN w/ (RGB, RGB) & $22.78$ & $34.81$ & $28.55$ & $52.14$ & $10.28$ & $31.25$ & $13.35$ & $42.20$ & $14.53$ & $32.96$ & $17.90$ & $69.85$ & $38.67$ \\
			TCMN w/ (RGB, Flow) & $21.96$ & $35.26$ & $34.59$ & $61.66$ & $14.93$ & $37.52$ & $18.25$ & $48.87$ & $13.46$ & $39.42$ & $20.64$ & $71.77$ & $44.55$ \\
			TCMN w/ (Flow, RGB) & $27.03$ & $40.86$ & $29.81$ & $56.55$ & $16.58$ & $37.03$ & $16.62$ & $46.70$ & $15.44$ & $38.60$ & $21.10$ & $73.13$ & $43.94$ \\
			TCMN w/ (Flow, Flow) & $27.03$ & $39.55$ & $33.71$ & $58.08$ & $20.56$ & $43.43$ & $20.47$ & $50.84$ & $19.88$ & $40.04$ & $\mathbf{24.33}$ & $76.65$ & $46.39$ \\
			\hline
			Ensemble TCMN (Small) & $28.00$ & $41.80$ & $\mathbf{36.44}$ & $58.35$ & $21.52$ & $43.54$ & $16.67$ & $46.57$ & $15.82$ & $37.61$ & $23.69$ & $76.02$ & $45.57$ \\
			Ensemble TCMN (Full) & $28.77$ & $42.37$ & $35.47$ & $\mathbf{59.28}$ & $17.91$ & $40.79$ & $\mathbf{20.47}$ & $\mathbf{50.78}$ & $18.81$ & $\mathbf{42.95}$ & $24.29$ & $\mathbf{76.98}$ & $\mathbf{47.24}$ \\
			\hline
		\end{tabular}
	\end{center}
	\caption{Ablation study on the effects of TCMN components on TEMPO-HL~\cite{hendricks17iccv}}
	\label{tab:ablation}
\vspace{-6mm}
\end{table*}

We examine several variants of our model to investigate the effect of each module in the model.

\subsubsection{The Effects of Different Modules}

In order to investigate the contribution of each module to the final result, we compare the model purely based on localization module without any attention, ``TCMN w/o Any Att'', TCMN without the attention in the localization module, ``TCMN w/o Loc Att'', and our full model, ``Ensemble TCMN (Full)''.
In Table~\ref{tab:ablation}, we observe that there is a large improvement after employing the relationship module and a relative small improvement after employing visual attention in the localization module.
Consequently, we show that both modules benefit the overall performance.

\subsubsection{Comparison between Different Streams}

When comparing the performance of each single stream, the model with the $(Flow, Flow)$ input, which is the method ``TCMN w/ (Flow, Flow)'' in Table~\ref{tab:ablation}, achieves the best average performance compared to the others.
The reason for the superior performance of $(Flow, Flow)$ is that a large portion of queries requires understanding of motion (including camera motion), such as \textit{the toy keeps spinning after it is originally spun} and \textit{the camera zooms in on the stick then the camera pans away}.

We also validate the effectiveness of the ensemble on different combinations of visual inputs.
As shown in Table~\ref{tab:ablation}, our ensemble model ``Ensemble TCMN (Full)'', which additionally introduces two streams $(RGB,Flow)$ and $(Flow,RGB)$, achieves better performance compared to the model ``TCMN w/ Two Streams'' that only utilizes two streams $(RGB, RGB)$ and $(Flow, Flow)$, which is the case in MLLC~\cite{hendricks18emnlp}. 
The performance of MLLC~\cite{hendricks18emnlp} can also be improved with two additional streams, as shown in ``Ensemble MLLC+conTEF''.
Therefore, we conclude that the four combinations of visual inputs are complementary for this task.

\subsubsection{The Number of Parameters}

In order to verify whether the improvement is caused by using more parameters, we shrink the hidden state size, and refer to the resulting model as ``Ensemble TCMN (Small)''.
``MLLC+conTEF'' has 17M parameters in their experimental settings, while ``Ensemble TCMN (Small)'' has 16.5M parameters.
We can see that even with fewer parameters, our method ``Ensemble TCMN (Small)'' still outperforms the state-of-art method ``MLLC+conTEF''.
Increasing the number of parameters can make further improvement, as shown in  ``Ensemble TCMN (Full)''

\subsubsection{Comparison with the Attention Network for Referential Expression Comprehension}
\label{sec:tree-vs-seq}

Another interesting question is whether introducing an external language parser is beneficial for decomposing the descriptions of events and temporal signals.
In order to answer this question, we compare our tree attention network with the sequential attention network widely used in referential expression comprehension~\cite{hu2017modeling,zhang2018grounding,yu2018mattnet,liu2019improving},  where a sequential LSTM is adopted as the  backbone.

In Table~\ref{tab:ablation}, the method ``Model w/ REC Att'' is the model adopting their sequential attention. 
We observe that such an adaptation outperforms ``MLLC+conTEF'' in all the metrics.
One potential reason is that softmax tends to encourage one value to be high and suppress others.
For simple sentences, the effectiveness of attending the single word based on sequential LSTM has been validated in many previous works~\cite{chen2018temporally,liu2018tmn,liu2018crossmodal,liu2018attentive,chen2019localizing}.
For complex sentences, since the evaluated temporal signals are all single words, this leads to the effectiveness of attention on sequential LSTM. 
Compared to our method, the performance of such adaptation on complex sentences is worse than ours, while the performance on simple sentences is better.
One possible reason that the tree attention network outperforms the sequential attention network on complex sentences is that it encodes structure information.
However, such structural encoding may not be detailed enough when handling simple sentences,  unlike the sequential attention network.



\section{Conclusions}
\label{sec:conclusion}

In this work, we introduce a temporal compositional modular network that first learns to softly decompose a sentence into three descriptions with respect to the main event, context event and  temporal signal, 
and then matches the cross-modal features with two separate modules in terms of visual similarity and location similarity between each segment and the decomposed descriptions.
Furthermore, an ensemble method that considers the deficiencies of different visual modalities makes additional improvement.
Experimental results on TEMPO demonstrate the effectiveness of our method when handling  temporal language. 
In the future, we plan to incorporate detailed interactions between words and frames that can benefit both simple and complex sentences. 
We would also like to make further exploration in social videos.

\section{Acknowledgments}

We thank the support of NSF awards IIS-1704337, IIS-1722847, IIS-1813709, and the generous gift from our corporate sponsors.

%
\bibliographystyle{ACM-Reference-Format}
\bibliography{reference}

\end{document}